# Understanding the origins of the intrinsic dead-layer effect in nanocapacitors

M. S. Majdoub[1], R. Maranganti[1], P. Sharma[1,2,♣]
[1]Department of Mechanical Engineering,
[2]Department of Physics
University of Houston, Houston, TX, 77204, U.S.A

**Abstract:** Thin films of high-permittivity dielectrics are considered ideal candidates for realizing high charge density nanoscale capacitors for use in next generation energy storage and nanoelectronics applications. The experimentally observed capacitance of such film nanocapacitors is, however, an order of magnitude lower than expected. This dramatic drop in capacitance is attributed to the so-called "dead layer" – a low-permittivity layer at the metal-dielectric interface in series with the high-permittivity dielectric. Recent evidence suggests that this effect is intrinsic in the sense that its emergence is evident even in "perfectly" fabricated structures. The exact nature of the intrinsic dead-layer and the reasons for its origin still remain somewhat unclear. Based on insights gained from recently published *ab initio* work on $SrRuO_3/SrTiO_3/SrRuO_3$ and our first principle simulations on Au/MgO/Au and Pt/MgO/Pt nanocapacitors, we construct an analytical model that isolates the contributions of various physical mechanisms to the intrinsic dead layer. In particular we argue that strain-gradients automatically arise in very thin films even in complete absence of external strain inducers and, due to flexoelectric coupling, are dominant contributors to the dead layer effect. Our theoretical results compare well with existing, as well as our own, *ab initio* calculations and suggest that inclusion of flexoelectricity is essential for qualitative reconciliation of atomistic results. Our results also hint at some novel remedies for mitigating the dead layer effect.

## I. Introduction

Next generation advances in energy storage and nanoelectronics require capacitors fabricated at the nanoscale. High dielectric constant materials such as perovskite ferroelectric materials are important candidates for such applications. Consider the following: the expected capacitance (based on classical electrostatics) of a 2.7 nm $SrTiO_3$ (STO) thin film is ~ 1600 fF$\mu$m$^{-2}$. *Ab initio* simulations on a $SrRuO_3/SrTiO_3/SrRuO_3$ (SRO/STO/SRO) capacitor system with the same dimensions however predict a much lower value --- 258 fF$\mu$m$^{-2}$! This dramatic drop in capacitance is traditionally attributed to the so-called "dead layer" effect[1]. The dead layer is thought to be as a low-permittivity thin layer at the metal/dielectric interface connected in series with the rest of the dielectric. Indeed early experimental work by Mead[2,3] and subsequently many others[4-7] have documented the effects of the disruptive dead-layer for nanometer sized films. The presence of this dead layer is attributed to a variety of reasons including a secondary low-permittivity phase at the surface of the films, nearby-surface variation of polarization (field induced or spontaneous)[8], presence of misfit dislocations[9,10], electric field penetration into the metal electrodes[11-14] among others. The effect of electric field penetration into the metal electrodes to explain the anomalous capacitance was proposed by Mead[2] himself and has since then been theoretically investigated by several authors[11-14]. Recent pioneering *ab initio* calculations by Stengel and Spaldin[15] on $SrRuO_3/SrTiO_3/SrRuO_3$ and $Pt/SrTiO_3/Pt$ thin film capacitors with atomistically smooth interfaces (to exclude effects due to, say, misfit dislocations) have confirmed the intrinsic nature of this effect and that electric field penetration occurs in real metal electrodes giving rise to a passive *dead layer* at the metal-dielectric interface.

---

[♣] Corresponding author: psharma@uh.edu

The effective capacitance $C_{eff}$ of such a system with regions of low interfacial capacitance density at the metal-dielectric layer (i.e. the dead layer) $C_i$, connected in series with the nominal capacitance $C_0$ of the dielectric is typically expressed as:

$$\frac{1}{C_{eff}} = \frac{1}{C_i} + \frac{1}{C_0} + \frac{1}{C_i} \tag{1}$$

The interfacial capacitance $C_i$ is taken as the additional capacitance introduced into the system due to the penetration of the electric field into the metal electrodes, while the nominal capacitance $C_0$ of the dielectric layer is that predicted by classical electrostatics: for example, for a parallel plate capacitor made up of a dielectric with dielectric permittivity $\varepsilon$ and thickness $d$:

$$C_0 = \frac{\varepsilon}{d} \tag{2}$$

The calculations performed by Stengel and Spaldin[15] provide a much deeper understanding of the origins of the dead-layer. Indeed, their results, exemplified on the SRO/STO/SRO capacitor system show that the physical picture painted by Equations (1) and (2) may be incorrect and certainly incomplete. In particular, their results show that the electrostatic potential profile in the dielectric part of the capacitor exhibits considerable non-linear behavior (as opposed to the linear variation predicted by classical electrostatics) and that the capacitance of the dielectric layer $C_0$ is subject to some additional size-dependent scaling beyond what is suggested by Equation (2) alone. We agree with the results of the *ab initio* work of Spaldin and Stengel[15], however, in order to extract additional insights into the underpinning of the dead-layer; we construct a theoretical model of the possible underlying physical mechanisms and carry out our own *ab intio* calculations for other nanocapacitor systems. In this work, we argue that electric field penetration occurs inside the metal electrodes due to the diffuse nature of the metal-dielectric interface which in-turn triggers a *secondary* mechanism wherein intrinsic strain-gradients arise in thin films activating the flexoelectric effect (strain gradient-induced polarization). Ironically, the *secondary* mechanism is found to dominate! Furthermore, our analytical approach, allows a facile means to infer the correct scaling behavior of thin film capacitance (something that is beyond the computational power of purely *ab initio* based simulations since the latter calculations are limited to a few nanometers thick films).

We note that the effect of strain-gradients *induced due to lattice mismatch* in $Ba_{0.5}Sr_{0.5}TiO_3$ (BST) dielectric films grown on $SrRuO_3$ (SRO) metal electrodes has already been investigated by Catalan *et al.*[16] who predict that the large flexoelectric coefficients measured by Cross *et al.*[17] and Zubko *et al.*[18] for ferroelectrics like BST can potentially result in considerable changes in the polarization and permittivity behavior of thin film ferroelectric capacitors. Since the *ab initio* simulations are performed in the absence of lattice mismatch effects, we shall ignore the presence of such *extrinsic* strains in the present work.

As mentioned previously, we postulate that the two dominant mechanisms responsible for polarization effects which lead to substantial decrease in the capacitance of thin dielectric-metal electrode capacitor systems are electric field penetration in metals and flexoelectricity in dielectrics respectively. While the former is well-evident from the *ab initio* results of Stengel and Spaldin, the latter will be justified in due course by its ability to correctly (albeit qualitatively) predict the basics physics behind the dead layer. With appropriate information from *ab initio* results, quantitative agreement is also achieved. As will be shown in this work, failure to invoke flexoelectricity (and reliance only on electric field penetration in the metal as the mechanism behind intrinsic dead-layer) cannot reconcile the *ab initio* results (both ours as well as Reference[15] calculations).

The outline of this paper is as follows. In Section 2, after a brief summary of the pertinent concepts related to the flexoelectric phenomena, we solve the problem of a simple thin film based metal-insulator-metal capacitor system and highlight the relevance of flexoelectricity. The role of electric field penetration in metal electrodes is discussed in Section 3 and related to our central (flexoelectricity-based) results in Section 2. We present the results of our model in Section 4 with specific application to SRO/STO/SRO based nanocapacitor system and draw a comparison with existing *ab initio* results of Stengel and Spaldin[15]. In this section, to elucidate the physical insights as well provide further prove of our conjectures, we also present *ab initio* calculations on other materials systems (Au/MgO/Au and Pt/MgO/Pt)

**II. Flexoelectricity and Consequences for Dead-Layer in Nanocapacitors**

In the traditional continuum field theory of piezoelectric materials, an electric polarization is generated in response to uniform strain (or vice versa). Within the assumptions of linearity, a third-rank piezoelectric tensor **d** relates the polarization vector **P** to the second-rank strain tensor **S**,

$$(\mathbf{P})_i = (\mathbf{d})_{ijk}(\mathbf{S})_{jk} \qquad (3)$$

**d** being a third-order tensor, symmetry considerations require that it vanish for materials possessing a center of inversion symmetry. However, under conditions of non-uniform strain, the inversion symmetry in centrosymmetric materials can be broken to induce a net polarization. Phenomenologically, Equation (3) can be extended to include the contribution of strain gradients:

$$(\mathbf{P})_i = (\mathbf{d})_{ijk}(\mathbf{S})_{jk} + (\boldsymbol{\mu})_{ijkl}\nabla_l(\mathbf{S})_{jk} \qquad (4)$$

This added effect is referred to as the flexoelectric effect[19-22] and the components of the fourth order tensor **μ** are called the flexoelectric coefficients. The reader is referred to previous works for a review of this phenomena e.g. [Tagantsev[21,23], Cross[24]] and more recently our works[25-27]. Kogan[19] suggested that $e/a$, is a suitable lower bound for the flexoelectric constants for crystalline dielectrics, where $e$ is the electronic charge and $a$ is lattice parameter. Others have suggested that multiplication by relative permittivity is more appropriate[28], which now appears to have been confirmed experimentally[17,18].

Therefore high dielectric constant insulators like non-piezoelectric paraelectric STO are expected to exhibit large flexoelectric effects in certain directions.

We note that Yang et al.[29,30] have investigated the use of non-local polarization law to study the size effects and electromechanical coupling in thin film nanocapacitors. Recently, Kalinin and Meunier[31] have investigated flexoelectricity in low-dimensional nanostructures. Our recent atomistic simulations and calculations on prototype nanostructures[26] have revealed a striking enhancement in the effective piezoelectric constant of nearly 500 % over bulk for tetragonal $BaTiO_3$ cantilever beam around 5 nm (and a corresponding 80 % increase for the non-piezoelectric cubic phase at the same size). In a more recent work[27], we have also explored the use of the flexoelectric effect in nanostructures for energy harvesting applications. Results show a dramatic enhancement in energy harvesting for a narrow range of dimensions in such piezoelectric nanostructures.

We have presented in previous works[26,32] a detailed mathematical theory of flexoelectricity. The governing equations valid for a dielectric occupying a volume V bounded by a surface S in a vacuum V' are:

$$\nabla.\boldsymbol{\sigma} + \mathbf{f} = 0 \text{ where } \boldsymbol{\sigma} = \mathbf{T} - \nabla.\tilde{\mathbf{T}} \text{ in V}$$
$$\bar{\mathbf{E}} + \nabla.\tilde{\mathbf{E}} - \nabla\varphi 0 = 0 \text{ in V} \qquad (5)$$
$$-\varepsilon_0 \Delta\varphi + \nabla.\mathbf{P} = 0 \text{ in V and } \Delta\varphi = 0 \text{ in V'}$$

**σ** may be considered as the actual physical stress experienced by a material point and differs from the Cauchy stress **T**. The remaining variables are defined through the following constitutive relations:

$$\mathbf{T} = \mathbf{c}:\mathbf{S} + \mathbf{e}:\nabla\mathbf{P} + \mathbf{d}.\mathbf{P}$$
$$\tilde{\mathbf{T}} = \mathbf{f}.\mathbf{P}$$
$$-\bar{\mathbf{E}} = \mathbf{a}.\mathbf{P} + \mathbf{g}:\nabla\mathbf{P} + \mathbf{f}:\nabla\nabla\mathbf{u} + \mathbf{d}:\mathbf{S} \qquad (6)$$
$$\tilde{\mathbf{E}} = \mathbf{b}:\nabla\mathbf{P} + \mathbf{e}:\mathbf{S} + \mathbf{g}.\mathbf{P}$$

The coefficients of the displacement, polarization and their gradients defined above as "**a**", "**b**", "**c**", "**d**", "**f**", "**g**" and "**e**" are material property tensors. The second order tensor "**a**" is the reciprocal dielectric susceptibility. The fourth order tensor "**b**" is the polarization gradient-polarization gradient coupling tensor and "**c**" is the elastic tensor. The fourth order tensor "**e**" corresponds to polarization gradient and strain coupling introduced by Mindlin[22] whereas "**f**" is the fourth order flexoelectric tensor. "**d**" and "**g**" are the third order piezoelectric tensor and the polarization-polarization gradient coupling tensor. Further details may be obtained via consultation of References[26,32].

The corresponding boundary conditions on S described by normal vector **n** are:

$$\boldsymbol{\sigma}.\mathbf{n} = \mathbf{t} \text{ where } \boldsymbol{\sigma} = \mathbf{T} - \nabla.\tilde{\mathbf{T}}$$
$$\tilde{\mathbf{E}}.\mathbf{n} = 0 \quad\quad\quad (7)$$
$$(-\varepsilon_0 \|\nabla\varphi\| + \mathbf{P}).\mathbf{n} = 0$$

The symbol $\|\ \|$ denotes the jump across the surface or an interface.

We analyze the dielectric response of a thin film capacitor system illustrated in Figure (1). The dielectric is considered to have a cubic *centrosymmetric* lattice[33]. The one-dimensional flexoelectric equations (varying along the *z*-direction) for the system shown in Figure (1) reduce to:

$$c_{11}\frac{\partial^2 u}{\partial z^2} + (e_{11} - f_{11})\frac{\partial^2 P}{\partial z^2} = 0$$
$$(e_{11} - f_{11})\frac{\partial^2 u}{\partial z^2} + b_{11}\frac{\partial^2 P}{\partial z^2} - a_{11}P - \frac{\partial \phi}{\partial z} = 0 \quad\quad (6\text{a-c})$$
$$-\varepsilon_0 \frac{\partial^2 \phi}{\partial z^2} + \frac{\partial P}{\partial z} = 0$$

The differential equations (8a-c) have to be solved subject to boundary conditions which we specify as follows. Firstly, the mechanical force must vanish at the metal-dielectric interface which gives us our first boundary condition as:

$$\left.(c_{11}\frac{\partial u}{\partial z} + (e_{11} - f_{11})\frac{\partial P}{\partial z})\right|_{z=\pm L} = 0 \quad\quad (7)$$

Also, due to the voltage drop in the metal due to electric field penetration, the actual potential $\pm V_d$ at the metal-dielectric interface will differ from the one applied $\pm V$ (See Figure (1)) and can be specified as

$$\left.\phi\right|_{z=\pm L} = \pm V_d \quad\quad (8)$$

$V_d$ remains to be determined.

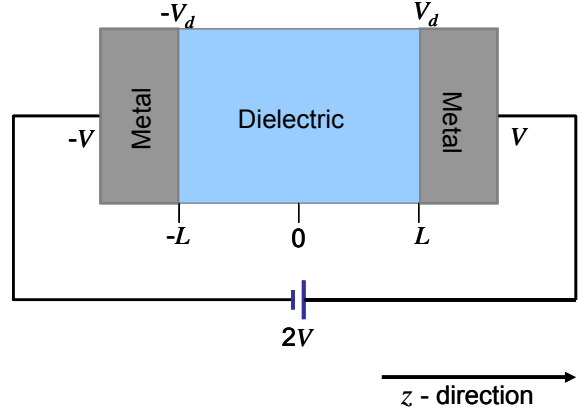

**Figure 1**: Schematic of a parallel plate capacitor

While, the boundary conditions given by Equations (9) and (10) would be enough to solve the system in Figure (1) in the absence of flexoelectricity (i.e. the case where $e_{11}=0$, $f_{11}=0$ and $b_{11}=0$), the presence of flexoelectricity requires the specification of an additional boundary condition, which may be taken as the specification of the polarization at the metal-dielectric interface by:

$$P\big|_{z=\pm L} = -k\varepsilon_0 \eta V_d / L \qquad (9)$$

$\eta = \dfrac{\varepsilon_d - \varepsilon_0}{\varepsilon_0}$ is the dielectric susceptibility where $\varepsilon_d$ is the dielectric constant of the dielectric and $k$ is a constant that controls the depth at which the electric field penetration occurs.

The solution to the three independent fields: displacement $u$, polarization $P$ and electric potential $\phi$ can be adapted from Mindlin[34] (who used a different physical theory, the so-called polarization gradient theory, but which has a mathematical structure similar to ours---our model contains his theory as well).

$$\begin{aligned} u &= B_1 \cosh(z/l) \\ P &= A_2 + B_2 \cosh(z/l) \\ \phi &= A_3 z + B_3 \sinh(z/l) \end{aligned} \qquad (10\text{a-c})$$

Here,

$$\begin{aligned} B_3 &= (1-k)\eta V_d / \left[\eta \sinh(L/l) + (L/l)\cosh(L/l)\right] \\ A_3 &= (B_3/\eta l)\cosh(L/l) + k V_d / L \\ A_2 &= -\varepsilon_0 \eta A_3,\ B_2 = \varepsilon_0 B_3 / l = -c_{11} B_1 / d_{11} \\ l &= \left[\varepsilon_0 \left(b_{11} c_{11} - (e_{11} - f_{11})^2\right) / c_{11}\left(1+\eta^{-1}\right)\right]^{-1/2} \end{aligned} \qquad (11\text{a-d})$$

Note that the flexoelectric constants $b_{11}$ and $e_{11}$ and $f_{11}$ occur in the solution for the fields only through a length parameter $l$ defined in Equation (13d). We will denote this length $l$ as the longitudinal flexoelectric length scale. As claimed earlier, the solution to the governing equations of flexoelectricity confirm that strain-gradients are automatically induced despite the absence of any external strain sources.

The position dependent "apparent" permittivity of the dielectric can be then written as:

$$\varepsilon(z) = \frac{\varepsilon_0 (1+\eta) A_3}{A_3 + (B_3/l)\cosh(z/l)} \tag{12}$$

Finally, the capacitance of the dielectric layer $C_d$ can be found out as the ratio between the electric displacement to the voltage across the layer as:

$$C_d = \frac{(\varepsilon_0 \partial \phi - P)_{z=\pm L}}{2V_d} = \frac{\varepsilon_0(1+\eta)}{2L} \frac{1+(k\eta l/L)\tanh(L/l)}{1+(\eta l/L)\tanh(L/l)} \tag{13}$$

As manifest from Equation (15), for "large" thicknesses, the capacitance reverts to that predicted by classical electrostatics.

### III. Electric Field Penetration

Now, we focus our attention on the remaining constituent of the capacitor i.e. the metal electrode. In the conventional picture of a capacitor, the electrode is an ideal metal and the electrical field inside the dielectric is perfectly screened. However, both experiments and *ab initio* simulations have shown that in real systems, screening of electric fields takes place over a finite spatial extent inside the metal. Because of the penetration of electric field into the metal, there is a potential drop inside the metal electrode which then introduces an additional capacitance into the system (apart from that due to the dielectric--Equation (15)). Typically, this effect is modeled by requiring the free charges in the electrode to form a layer of finite thickness at the metal-dielectric interface. In the conventional picture of a capacitor, the free charges reside at an infinitesimally thin layer at the metal-dielectric interface as a delta function and there is no separation between them and the polarization bound charge in the dielectric. However, when a free charge layer of finite thickness is assumed in the electrode, the center of charge in the electrode is separated by a finite distance from the polarization bound charge in the dielectric and an additional capacitance is introduced. Also, as one can infer from the electrostatic Poisson's equation, a finite spatial distribution of charges inside the electrode results in electric fields penetrating into the metal electrode: *a scenario which is forbidden in the conventional description of a capacitor*.

Dawber and Scott [12] have provided a description of the phenomenon of electric field penetration into metal electrodes based on the Drude model for a free electron gas in contact with a dielectric and arrived at physically insightful analytical expressions for the charge density distribution $\rho(z)$ and the electric field $E(z)$ inside the electrodes:

$$\rho(z) = \frac{Q}{\lambda}\exp\left(\frac{-|z|+L}{\lambda}\right), \quad E(z) = -\frac{Q}{\varepsilon_e}\exp\left(\frac{-|z|+L}{\lambda}\right) \tag{14}$$

$\lambda$ is the Thomas-Fermi screening length and is related to the static conductivity $\sigma_0$, diffusion coefficient $D$ and the dielectric constant of the metal electrode $\varepsilon_e$ as $\lambda = (\frac{4\pi\sigma_0}{D\varepsilon_e})^{-\frac{1}{2}}$. From the potential distribution, the corresponding capacitance of the metal electrode $C_e$ is $\frac{\varepsilon_e}{\lambda}$. Since the potential drop is supposed to happen entirely in an electron-gas like medium, the dielectric constant $\varepsilon_e$ is typically taken as that of free space. For an SRO electrode in the system under consideration, with $\lambda$ =0.5 Å[12], the value of this capacitance is $\approx 177\text{fF}/\mu\text{m}^2$; the net contribution to the capacitance due to both the electrodes is therefore around $88\text{ fF}/\mu\text{m}^2$. Hence, the capacitance of the whole system (after including the contribution due to the dielectric layer) cannot be larger than $88\text{ fF}/\mu\text{m}^2$. However, the capacitance of the SRO/STO/SRO system as found by Stengel and Spaldin[15] is ~ $258\text{fF}/\mu\text{m}^2$. Clearly, this approach overestimates the capacitance contribution due to the metal electrodes. A major criticism to this approach is that the capacitance contribution to metal electrode is independent of the size and the material properties of the dielectric region. It should also be noted that yet another approach to model the electric field penetration into metal electrodes involves consideration of the band structure of metal/dielectric interface[35]. However, the capacitance contributions of the metal electrodes using this approach are again overestimated. In order to address this issue, we propose an alternate approach to model the penetration of electric fields into the metal electrodes.

The idea of an abrupt metal/dielectric interface, especially while modeling phenomena varying at the level of a few angstroms is questionable since one would expect the atoms at the interface to have bonding that is intermediate in nature to that in the metal electrode and the dielectric. In order to model the diffuse interface, we will let the dielectric permittivity at the interface to be a continuously varying function of position[36]. Though the interface extends both into the metal and the dielectric, we will only subject the dielectric permittivity of the metal to the variation. In order to do so, we choose that the inverse dielectric permittivity of the metal $\varepsilon_e(z)^{-1}$ has the following functional form:

$$\frac{1}{\varepsilon_e(z)} = A\exp\left(-\frac{|z+L|}{l_{tr}}\right) \tag{15}$$

The constant $A$ is the inverse permittivity of the dielectric at the metal/dielectric interface which can be obtained by substituting $z=L$ in Equation (14) and $l_{tr}$ is the length-scale associated with the variation: the smaller the parameter $l_{tr}$, the sharper is the variation in the dielectric constant in the metal. Equations (14) and (17) provide the dielectric permittivity profiles in the dielectric and the metal respectively.

## IV. Theoretical Results, First Principles Calculations and Physical Insights

In this section we apply the preceding analysis to a SRO/STO/SRO capacitor system which has been investigated using *ab initio* methods by Stengel and Spaldin[15]. The dimensions of the capacitor (2$L$ =2.7nm), voltage applied (2$V$ = 27.8 mV) and the bulk permittivity of the dielectric STO (~490) are the same as the ones used in Reference[15]. The flexoelectric coefficients for STO have been determined by employing Askar *et al.*'s[37,38] approach and the longitudinal flexoelectric length scale is found to be 1.51 Å. However it is found that the transverse flexoelectric length scale is as large as ~3nm. The relatively large transverse value is the reason behind observation of large flexoelectricity effects only in shearing or bending experiments[17,18] but does not enter the one-dimensional equations for our particular problem. The constant $k$ is estimated from the average dielectric permittivity of the system in the *ab initio* calculations of Reference[15] as 0.3 and the transition length scale is 2.0 Å. With these parameters and use of Equation (15), we are able to find a good quantitative agreement with the *ab initio* results of Reference[15] : 256 fF/μm$^2$. The major insights however lie in the qualitative comparison and the scaling of capacitance which we now proceed to make.

In Figure (2), we plot the scaling of capacitance with size as predicted by (i) classical electrostatics, (ii) if only the flexoelectricity mechanism is operative, (iii) if only electric field penetration is operative and finally (iv) if both flexoelectricity and electric field penetration are operative. The lone data point available from *ab initio* calculations of Reference[15] is also shown. Results clearly illustrate that neither flexoelectricity alone nor electric field penetration alone can reconcile the *ab initio* results correctly and both are (not only present) but needed. To further appreciate the latter statement, we note that if only electric field penetration is considered as the mechanism, the scaling of capacitance with size cannot be reconciled. Flexoelectricity however is insufficient (by itself) to quantitatively match the *ab initio* results.

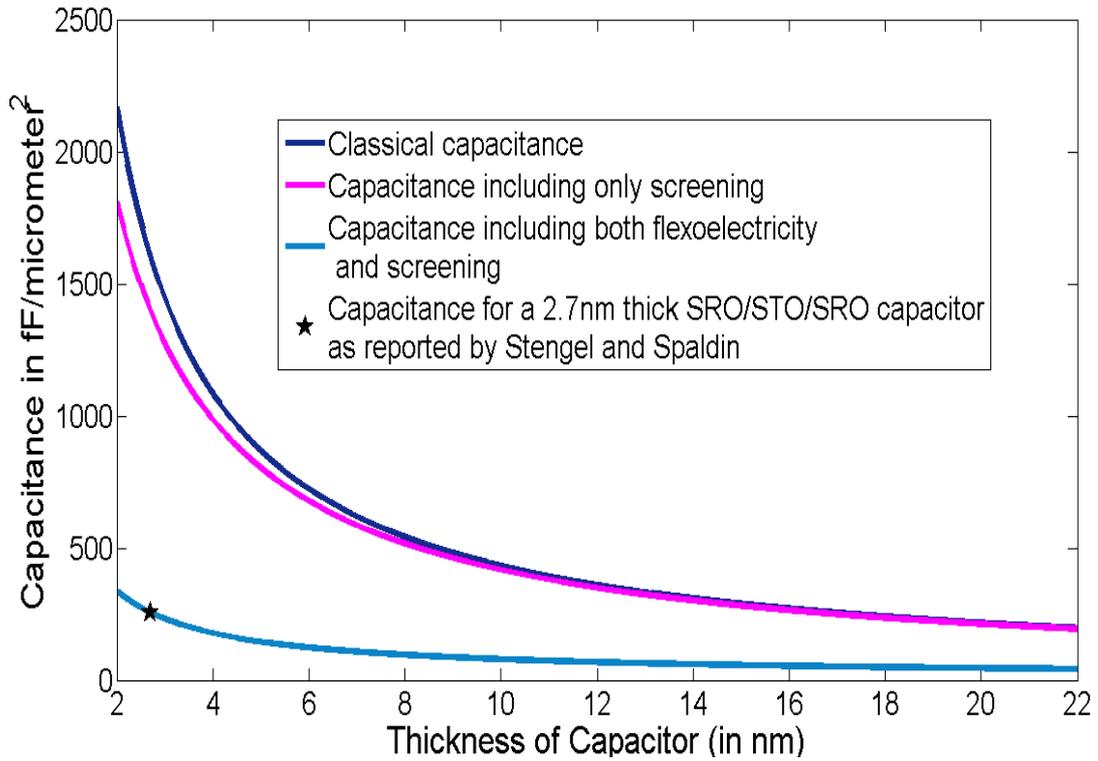

**Figure 2:** Thickness size dependent capacitance of SRO/STO/SRO capacitor. Only if both screening and flexoelectricity are considered we get a good agreement with *ab initio* calculations.

The voltage at the metal/dielectric interface 6.80 mV and the electric field in the middle of the dielectric (1.6 Vµm$^{-1}$) compare favorably with the *ab initio* results. The profile of electrostatic potential (Figure 3) also exhibits a reasonable match. We note that the degree of non-linearity of the induced potential predicted by the continuum flexoelectric theory (as opposed to the linear behavior predicted by classical electrostatics) depends crucially upon the size of the thin film and the bulk dielectric permittivity of the constituent insulating material.

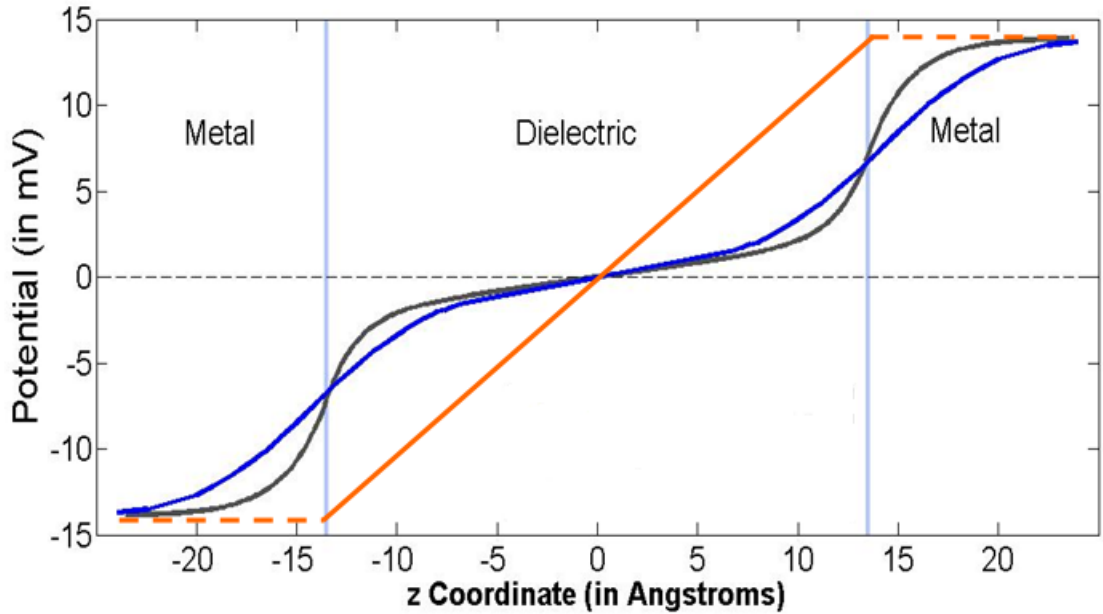

**Figure 3:** Comparison of the potential profile inside the SRO/STO/SRO capacitor system obtained by continuum flexoelectricity (blue curve) with that obtained by Stengel and Spaldin[15] using *ab initio* techniques (black curve). Red curve is the potential expected from classical electrostatics.

For a bulk sized-film, the non-linearity in the potential profile predicted by flexoelectricity confines itself to a very small region near the interface: the resulting change in capacitance is therefore negligible and hence this effect is of little or no consequence for such bulk-sized films. Further, this non-linearity also depends directly upon the dielectric permittivity of the material: i.e. the better the dielectric, the larger is the non-linearity in the induced potential and the consequent adverse impact on the capacitance. Therefore it is expected that nanometer sized thin-film capacitors made of high-permittivity materials such as ferroelectric perovskites will be the most affected by the adverse impact of flexoelectricity. Indeed *ab initio* simulations on nanometer sized Pt/MgO/Pt systems[15] (dielectric constant $\varepsilon$ of MgO is 9.8) show virtually no anomalous capacitance behavior. In the remaining of this work, we try to investigate further the relevance of this conclusion using first principles calculations on Au/MgO/Au and Pt/MgO/Pt nanocapacitors. We also point out the challenges beyond such calculations.

Most of the finite electric field calculations within the framework of Density Functional Theory (DFT) are based on the Berry phase formalism. However, in the case of realistic nanocapacitors with a metal-dielectric interface, the requirement of a unique Fermi level in DFT calculations is broken. In the presence of electric field, an effective bias potential is induced due to the shift in the Fermi levels creating a partially occupied states at the energy gap of the dielectric[39]. A possible method to overcome this difficulty with finite electric field is provided by the non-equilibrium Green's functions, but this method turns out to be not computationally efficient. Stengel and Spaldin[15,39] implemented an alternative method based on Wannier-function theory to correctly simulate the Metal/Insulator/Metal system. They also suggested another approach, which uses conventional *ab initio* codes such as VASP (Vienna *ab initio* Simulation Package) and PWscf (Plane Wave Self Consistent Field), by simulating only the Metal/Insulator in the presence of vacuum. This method was successfully implemented by Lee *et al.*[40] and used to calculate the capacitance of Au/MgO/Au and Ni/ZrO$_2$/Ni systems.

We used VASP[41] to study a system of Au/MgO(100)/Au and Pt/MgO(100)/Pt. Only the Metal/Insulator (MI) slab (half of the realistic capacitor is modeled) with both sides exposed to vacuum in the presence of external electric field (to model the voltage bias between the electrodes) is simulated (see Figure 4). A uniform electric field is applied along the z direction and is modeled by adding a saw-tooth like potential to the external potential entering the Kohn-Sham equations. We used the Projector Augmented Wave (PAW)[42] potentials to describe the ionic potentials and Local Density Approximation to represent the exchange correlation energies. A regular 6x6x1 k-points mesh and 400 eV energy cutoff were employed. A Gaussian smearing with width 0.5 eV was used to describe the partial occupancies for each wavefunction. The longitudinal lattice mismatch between gold and the oxide is within 1% whereas the transversal lattice constants considered are those of the oxide (in-plane lattice parameters were set to 4.15 nm). The system is then relaxed until Hellmann-Feynman forces on each atom are below 0.02 eV/A. Since the system under study has non-polar terminated surfaces, we do not need to treat or passivate the surfaces. The vacuum region in each supercell was large enough to avoid wavefunction overlaps and interactions between neighboring supercells. The plane-wave basis set used in most of the DFT calculations assumes a full periodicity of the supercell geometry and the electrostatic potential. In general, for surface supercell calculations in vacuum, the simulated slabs can be asymmetric which means that the periodic boundary condition is not satisfied since the electrostatic potential value will be different on the boundary of each neighboring supercells. Bengtsson[43] provided a solution to the potential mismatch at the boundary by introducing an artificial dipole correction to the energy expression for periodic supercell calculations. Such correction was implemented in VASP package and used to correctly estimate the electrostatic potential for supercell slab calculations in vacuum. A restriction to this technique is that the internal electric filed is limited by the permittivity of the dielectric. The internal electric field is equal to the external electric field in the vacuum divided by the dielectric constant. Hence, for high permittivity dielectric such as SrTiO$_3$, only small values of the electric fields are allowed. One more limitation is that the terminated surface of the insulator needs to be non-metallic otherwise we will recover the same situation in MIM system.

In order to obtain the total local potential and permittivity profiles, an averaging technique is used to smooth the variations. We consider a 0.1 eV/Å external electric field E$_{ext}$ applied along the axial z direction of the MI system. The ions are relaxed in response to the applied electric filed.

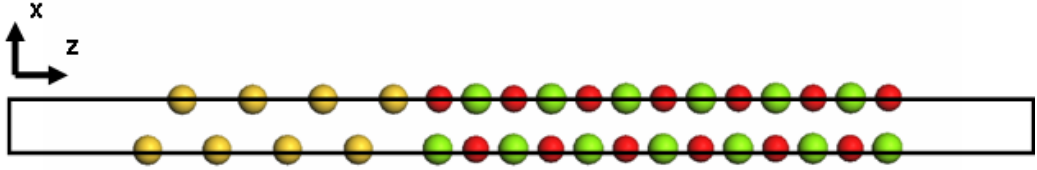

**Figure 4:** (Color online) Au/MgO simulated supercell. The metal (Au in yellow) is put into contact with the magnesium oxide (Mg in green and O in red) inside vacuum (right and left empty space).

The change in electrostatic potential $\Delta V$ is defined as the difference in electrostatic potentials V in the presence and in the absence of external electric field (averaged over the in plane xy cross section):

$$\Delta V(z) = \frac{1}{ab}\int_0^a \int_0^b [V(x,y,z)|_{E_{ext}} - V(x,y,z)|_{E_{ext}=0}]\, dydx \qquad (18)$$

The constants a and b respectively designate the unit cell lengths along the x and y directions.

To soften the variation of $\Delta V$, we take a macroscopic average over the bulk periodicity $l_1$ and $l_2$ respectively of the metal and insulator as defined by Reference[44]:

$$\langle \Delta V(z) \rangle = \frac{1}{l_1 l_2} \int_{z-\frac{l_1}{2}}^{z+\frac{l_1}{2}} \int_{z'-\frac{l_2}{2}}^{z'+\frac{l_2}{2}} \Delta V(z'')\, dz''dz' \qquad (19)$$

Hence, the local permittivity is simply expressed in terms of the external electric field and the local potential:

$$\langle \varepsilon(z) \rangle = -\frac{E_{ext}}{\frac{d}{dz}\langle \Delta V(z) \rangle} \qquad (20)$$

Figure 5 represents the variation of the local dielectric constant inside the metal Au and insulator MgO for both fixed and relaxed calculations. The static permittivity (Figure 5: dashed blue curve)) can be decomposed into ionic and electronic contributions and is evaluated by letting the ions to relax freely under applied bias. The electronic response corresponds to the so called optical permittivity Figure 5: solid red curve) that can be obtained by fixing the ions to their zero electric field equilibrium configurations.

Ideally, the dielectric constant in metals is infinite, but near the electrode-insulator interface there is an electric field penetration. This exactly corresponds to a zero inverse permittivity in the metal that increases as one approach closer to the metal-insulator interface---in good agreement with the simulated profile. The static and optical dielectric constants are estimated to be 9.52 and 2.78 compared to 9.65 and 3.36 of Reference[40].

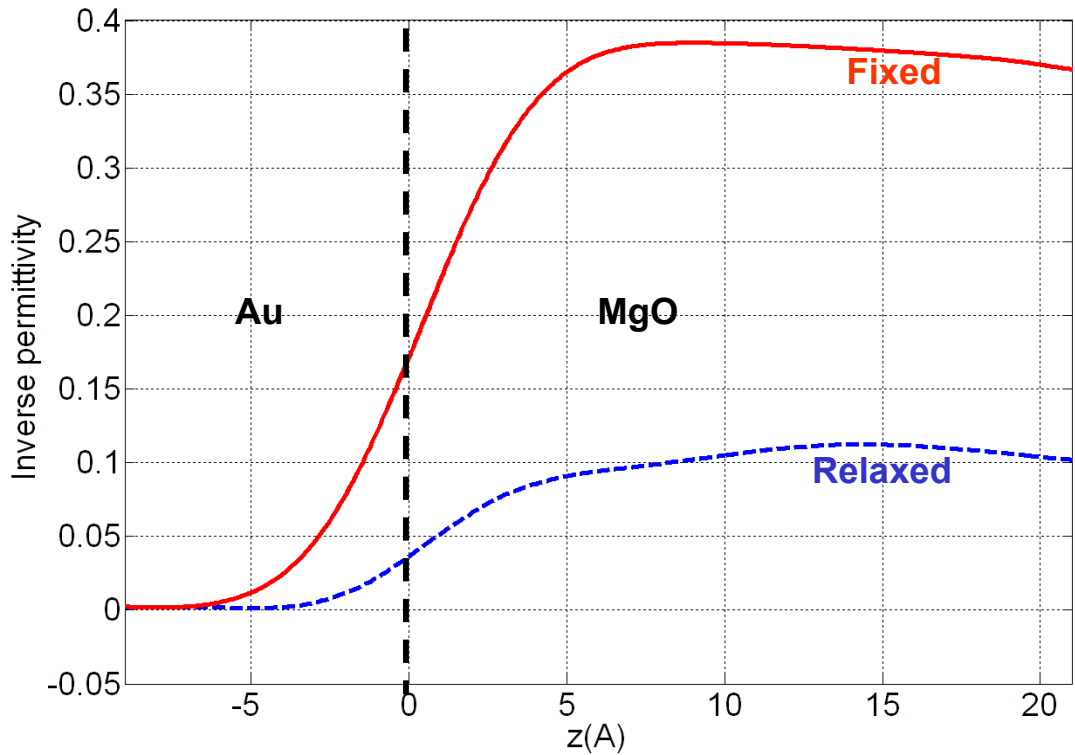

**Figure 5:** Au/MgO inverse local permittivity profile. The blue dashed line and solid red line are respectively for the relaxed and fixed ions simulations.

A comparison between our theoretical model (incorporating flexoelectricity and electric field penetration effects) and the *ab initio* calculation of the inverse permittivity profile is illustrated in Figure 6. Results show a good agreement between the theory and first-principle simulations. Curves tend to the bulk inverse static permittivity in the middle of the insulator. Our SRO/STO/SRO theoretical results match the *ab initio* simulations carried out by Stengel and Spaldin[15] for these nanocapacitor systems. Moreover, our Au/MgO/Au and Pt/MgO(100)/Pt calculations, show a good agreement and capture the right profile of the static permittivity with no anomalous behavior at the interface as expected for such low permittivity materials.

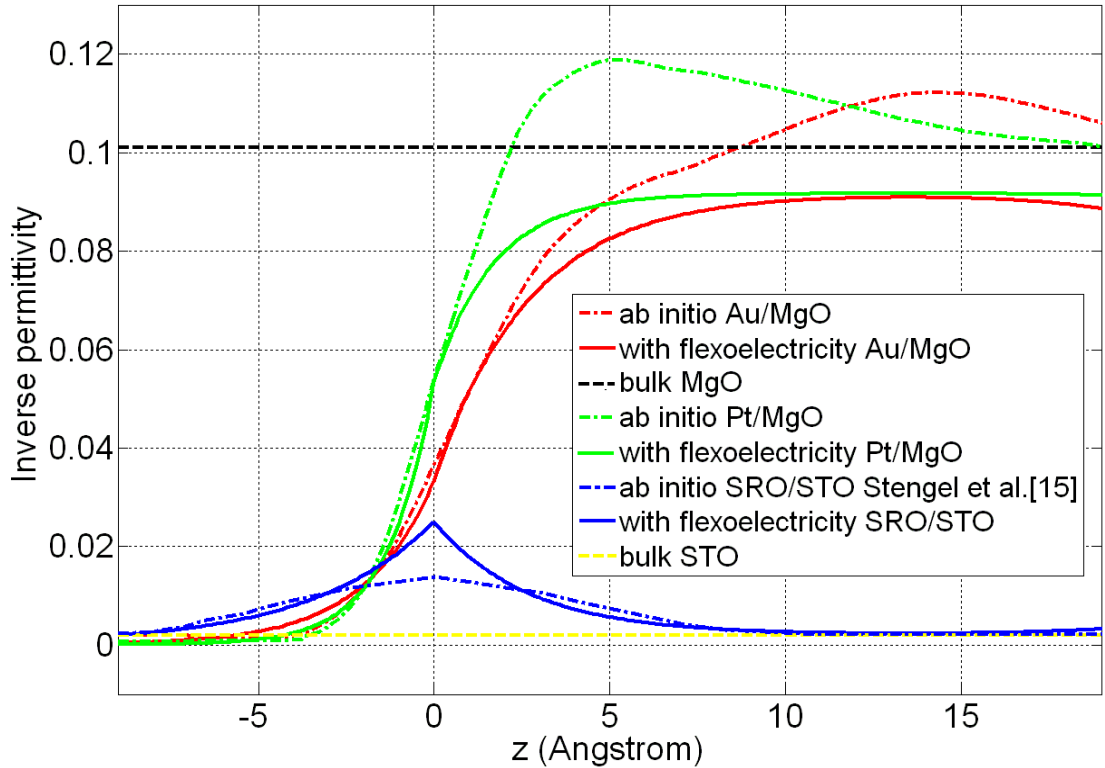

**Figure 6:** (color online) Inverse dielectric permittivity profiles for Au/MgO, Pt/MgO and SRO/STO. The horizontal dashed black and yellow lines are respectively the bulk MgO and STO inverse static permittivity. Results show a good agreement and match the bulk limit at the middle of the dielectric. Top left figure is a schematic of Au/MgO simulated supercell. The metal (Au in yellow) is put into contact with the magnesium oxide (Mg in green and O in red) inside vacuum (right and left empty space).

## V. Conclusion

Toward a better understanding of the phenomena beyond the dead layer at the electrode-dielectric interface, we have provided a theoretical model incorporating flexoelectricity combined with a new approach to electric field penetration inside metals that is able to explain well the first principles calculations performed by Stengel and Spaldin[15] and our own calculations carried out on Au/MgO/Au and Pt/MgO/Pt nanaocapacitors. Results indicate that both flexoelectricity and electric field penetration inside metal are essential to reconcile the atomistic simulations. Hence, flexoelectricity has an important contribution which opens the possibility of providing a remedy to the dead layer in nanocapacitors by carefully designing the metal-dielectric interface. Such an endeavor, based on the present results, will be taken in the future.

**Acknowledgements:**

Financial support from NSF NIRT Grant No. CMMI 0708096 is gratefully acknowledged.